\documentclass[aps,prl,reprint,groupedaddress,showpacs,amsmath,amssymb,twocolumn,floatfix]{revtex4-1}
\usepackage{graphicx}
\usepackage{bm}
\usepackage{color}
\usepackage[normalem]{ulem}
\usepackage{amsmath}
\usepackage{tabularx}

\usepackage[colorlinks=true]{hyperref}  
\hypersetup{
    unicode=false,          
    pdftoolbar=true,        
    pdfmenubar=true,        
    pdffitwindow=false,     
    pdfstartview={FitH},    
    pdftitle={XXX},    
    pdfauthor={XXX et al.},     
    pdfsubject={},   
    pdfcreator={},   
    pdfproducer={}, 
    pdfkeywords={} {} {}, 
    pdfnewwindow=true,      
    colorlinks=true,       
    linkcolor=magenta, 
    citecolor=blue,        
    filecolor=magenta,      
    urlcolor=blue           
} 
\usepackage{cleveref}

\newcommand{\Rpara}{$R_{\parallel}$}
\newcommand{\Rperp}{$R_{\perp}$}

\newcommand{\DRR}{$\Delta R / R_0$}

\usepackage{pifont}

\usepackage{bbm}

\renewcommand{\vec}[1]{\boldsymbol{#1}}

\begin{document}

\title{Evidence of Momentum Space Condensation in Rhombohedral Hexalayer Graphene}

\author{Erin Morissette$^{1}$}
\thanks{These authors contributed equally to this work.}
\author{Peiyu Qin$^{2}$}
\thanks{These authors contributed equally to this work.}
\author{K. Watanabe$^{3}$}
\author{T. Taniguchi$^{4}$}
\author{J.I.A. Li$^{1,2}$}
\email{jia.li@austin.utexas.edu}

\affiliation{$^{1}$Department of Physics, Brown University, Providence, RI 02912, USA}
\affiliation{$^{2}$Department of Physics, University of Texas at Austin, Austin, TX 78712, USA}
\affiliation{$^{4}$Research Center for Functional Materials, National Institute for Materials Science, 1-1 Namiki, Tsukuba 305-0044, Japan}
\affiliation{$^{4}$International Center for Materials Nanoarchitectonics,
National Institute for Materials Science,  1-1 Namiki, Tsukuba 305-0044, Japan}

\date{\today}

\maketitle

\textbf{
Spontaneous symmetry breaking provides a powerful window into the nature of underlying electronic orders. In strongly correlated systems, multiple symmetry-breaking orders can arise simultaneously. and their interplay generates an intricate landscape of quantum phases that has remained a central focus of condensed-matter research. In this work, we report a previously unidentified electronic phase in rhombohedral hexalayer graphene, distinguished by the simultaneous breaking of rotational, time-reversal, and inversion symmetries. Broken rotational symmetry is evidenced through anisotropic transport in angle-resolved measurements, while the onset of both the anomalous Hall effect and the nonlinear Hall effect signals the breaking of time-reversal and inversion symmetries. 
These combined signatures reveal an emergent order consistent with momentum-space condensation, a theoretically anticipated phenomenon realized here experimentally for the first time. This mechanism establishes a natural framework for understanding a broader class of correlated phases known to emerge from the flat bands of two-dimensional materials. 
}

\begin{figure*}
\includegraphics[width=1\linewidth]{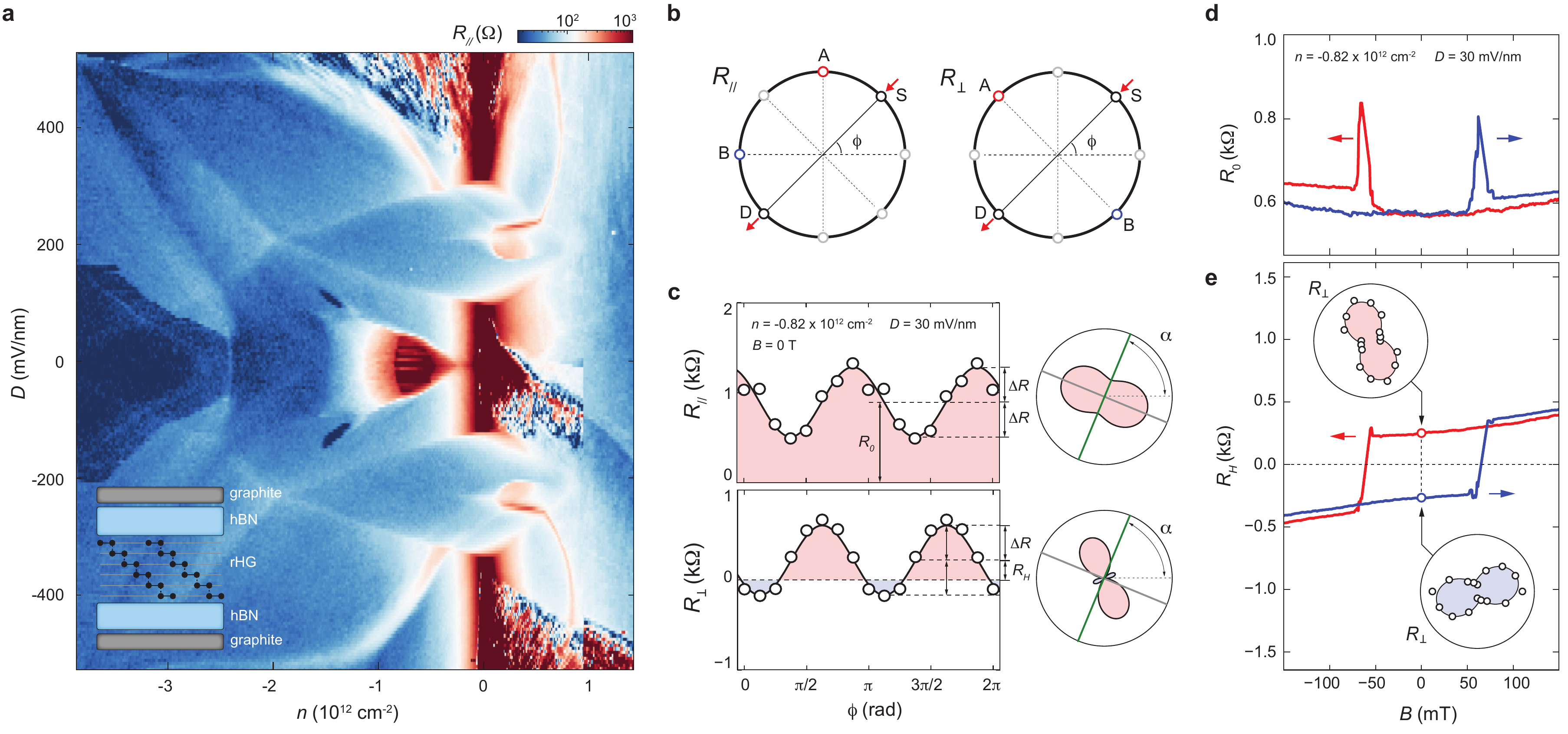}
\caption{\label{fig1}{\bf{Rhombohedral-stacking hexalayer graphene and angle-resolved transport measurement.}} 
(a) Longitudinal resistance \Rpara\ as a function of carrier density $n$ and displacement field $D$ in a dual-encapsulated, dual-gated rhombohedral hexalayer graphene device. The inset shows a schematic diagram of the heterostructure. 
(b) Schematic of the angle-resolved transport measurement setup used to extract $R_{\parallel}$ (left) and $R_{\perp}$ (right) in a disk-shaped sample.
(c) Angular dependence of $R_{\parallel}$ and $R_{\perp}$ inside the triangular multiferroic region, measured at $T = 40$ mK and $B = 0$ T. (d) $R_0$ and (e) $R_H$, extracted from the angular dependence using Eq.~1-2, measured as $B$ is swept back and forth. 
}
\end{figure*}

\begin{figure*}
\includegraphics[width=0.85\linewidth]{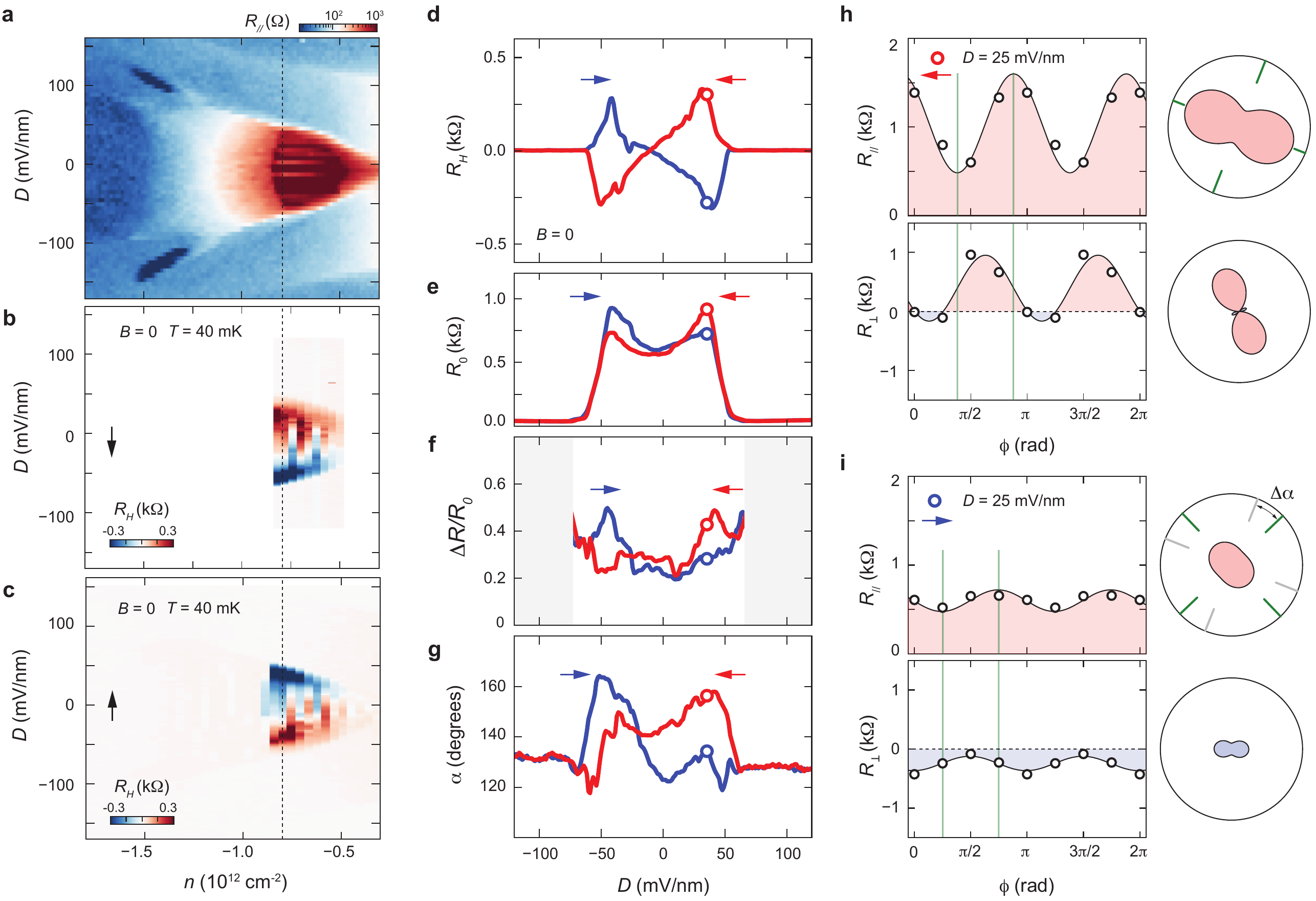}
\caption{\label{fig2}{\bf{Multiferroicity and electric-field-driven switching of transport anisotropy.}} Parameters describing the angular dependence are extracted from the multiferroic regime. 
(a) $R_0$ and (b-c) $R_H$ extracted from angle-resolved transport measurements as a function of $n$ and $D$ around the multiferroic regime. $R_H$ exhibits a sign reversal as $D$ is swept (b) from positive to negative, and (c) from negative to positive. As $D$ is swept back and forth at $n = -0.82 \times 10^{12}$ cm$^{-2}$, the multiferroic order manifests in the electric-field-driven switching in (d) the anomalous Hall effect $R_H$, which is accompanied by the hysteretic behaviors in (e) $R_0$, (f) \DRR, and (g) $\alpha$.  (h-i) Angular dependence measured at $n = -0.82 \times 10^{12}$ cm$^{-2}$ and $D = 25$ mV/nm as $D$ is swept (h) backwards and (i) forward.
}
\end{figure*}

\begin{figure*}
\includegraphics[width=1\linewidth]{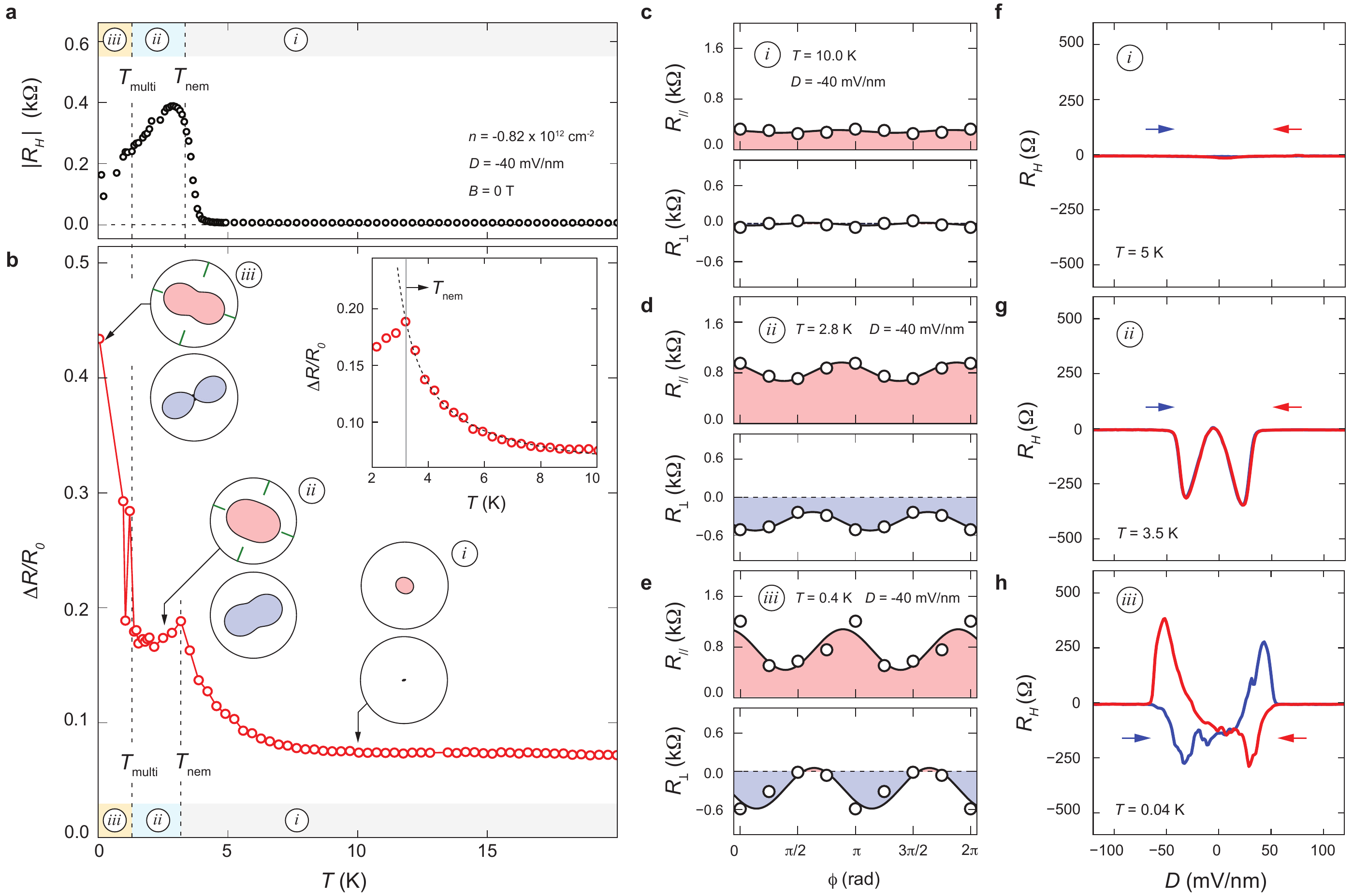}
\caption{\label{fig3}{\bf{Nematic phase transitions. }} (a-b) Temperature dependence of (a) the magnitude of the anomalous Hall coefficient $|R_H|$ and (b) the strength of transport anisotropy \DRR. Upper right inset shows a zoomed in plot of \DRR\ around $T_{nem}$, where the dashed curve denote the best fit using the Curie-Weiss law. (c-e) Angular dependence of \Rpara\ and \Rperp\ measured (c) in temperature regime i at $T = 10$ K, (d) in temperature regime ii at $T = 2.8$ K, and (e) in temperature regime iii at $T = 0.4$ K. (f-h) Anomalous Hall coeffecient $R_H$ measured with sweeping $D$ back and forth (f)) in temperature regime i at $T = 5$ K, (g) in temperature regime ii at $T = 3.5$ K, and (h) in temperature regime iii at $T = 40$ mK.     }
\end{figure*}

\begin{figure*}
\includegraphics[width=0.95\linewidth]{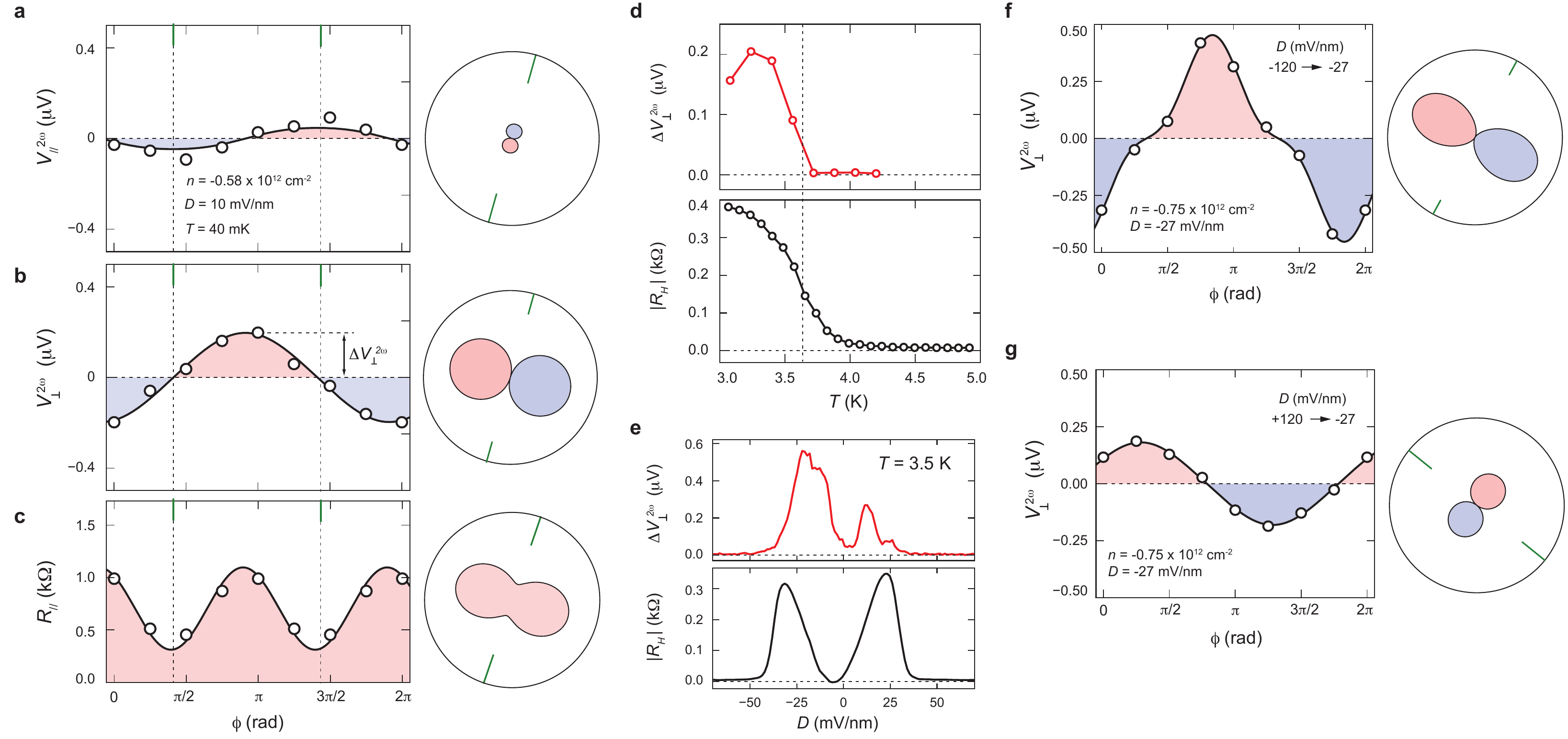}
\caption{\label{fig4}{\bf{Nonlinear Hall effect.}} Angular dependence of the second-harmonic nonlinear transport response (a) parallel ($V_{\parallel}^{2\omega}$) and (b) perpendicular ($V_{\perp}^{2\omega}$) to current flow, taken in the multiferroic regime. $\Delta V_{\perp}^{2\omega}$ denotes the angular oscillation in $V_{\perp}^{2\omega}$.  (c)  Angular dependence of linear transport response measured at the same $n$ and $D$ as panels (a) and (b). (d) $\Delta V_{\perp}^{2\omega}$ (top panel) and $|R_H|$ (bottom panel) measured as a function of $T$ at $D = 10$ mV/nm and $n=-0.58 \times 10^{12}$ cm$^{-2}$.  (e) $\Delta V_{\perp}^{2\omega}$ (top panel) and $|R_H|$ (bottom panel) measured as a function of $D$ at $T = 3.5$ K and $n=-0.58 \times 10^{12}$ cm$^{-2}$.  (f-g) The angular dependence of $V_{\perp}^{2\omega}$ measured at  $n=-0.75\times 10^{12}$ cm$^{-2}$ and $D = -27$ mV/nm, with different sweeping directions in $D$. 
}
\end{figure*}

Electronic phases that spontaneously break fundamental symmetries are ubiquitous in strongly correlated materials. Among these, rotational symmetry breaking, manifested as electronic nematicity, and time-reversal symmetry breaking, often associated with quantum magnetism, have emerged as central themes across a wide range of quantum materials. Recent theoretical developments have suggested that these two symmetry-breaking orders may be intrinsically intertwined in two-dimensional materials with flat energy bands~\cite{Jung2015momentum,Huang2023momentum,Dong2021momentum}. Driven by a distinctive Coulomb-induced flocking instability, charge carriers spontaneously condense into a single corner, or pocket, of the trigonally warped Fermi surface. The resulting electronic ground state acquires a finite net momentum, providing a natural mechanism that couples rotational and time-reversal symmetry breaking.

Because Coulomb interactions and trigonal warping provide the essential ingredients, spontaneous momentum condensation has been predicted in several graphene allotropes~\cite{Jung2015momentum,Huang2023momentum,Dong2021momentum}. In these systems, strong correlations associated with flat energy bands stabilize ordered phases that break either time-reversal symmetry~\cite{Serlin2019,Sharpe2019,Lin2021SOC,Lin2022SDE,Lu2024FCI,Choi2025R4G,Liu2024R4G,Sha2024R4G,Han2024R4G,Zhou2021RTG,Han2024RPG,Chen2020ABC,Zhou2021RTG,Han2023multiferroic} or rotational symmetry~\cite{Zhang2024nonreciprocity,Zhang2025angular,Chichinadze2024nonlinearHall,Cao2020nematicity}. Establishing whether these two broken-symmetry orders are intrinsically linked would provide a foundational framework for understanding the emergent electronic phases in these systems. 

However, direct identification of momentum polarization has remained elusive, owing largely to the experimental challenges associated with detecting the coexistence and intertwinement of rotational and time-reversal symmetry breaking using conventional probes~\cite{Fernandes2014nematicity,Bohmer2022nematicity,Ren2021NematicDecoupling,Calvera2024ValleyNematic}. A central open question therefore remains: what constitutes a definitive experimental signature of spontaneous momentum condensation?

In this work, we address this question through angle-resolved transport measurements in rhombohedral hexalayer graphene (R6G)~\cite{Wu2017nematic,Wu2020nematic,Vafek2023anisotropy,Chichinadze2024nonlinearHall,Zhang2024nonreciprocity,Zhang2025angular}. We show that momentum condensation leads to the simultaneous breaking of time-reversal, rotational, and inversion symmetries, manifested respectively through the emergence of orbital ferromagnetism, pronounced transport anisotropy, and the nonlinear Hall effect~\cite{Sodemann2015nonlinear,Ma2019nonlinear,Kang2019nonlinear,Chichinadze2024nonlinearHall}.

The low-temperature phase space of R6G is parameterized by the charge carrier density ($n$) and the displacement field ($D$). Transport properties across this parameter space are summarized in the color-scale map shown in Fig.~\ref{fig1}a. The near-perfect symmetry about $D = 0$ indicates the absence of substrate-induced sublattice symmetry breaking, consistent with R6G being misaligned from the encapsulation of hexagonal boron nitride. Our study focuses on the triangular region that emerges near $D = 0$ on the hole-doped side of charge neutrality, where a highly resistive response is represented by red in the chosen color scale.

To perform angle-resolved transport measurements~\cite{Chichinadze2024nonlinearHall,Zhang2024nonreciprocity}, the R6G sample is patterned into a ``sunflower'' geometry, consisting of a disk-shaped region with eight evenly spaced electrical contacts, as illustrated in Fig.~\ref{fig1}b. Transport measurements are carried out by applying a current bias between two contacts, designated as source (S) and drain (D), whose relative orientation defines the azimuthal direction of current flow ($\phi$). By recording the voltage response between contacts $A$ and $B$, we extract both the longitudinal (\Rpara) and transverse (\Rperp) resistances.

Fig.~\ref{fig1}c shows the angle-dependent transport response measured within the triangular regime. As the current direction $\phi$ is varied, both the longitudinal (\Rpara) and transverse (\Rperp) resistances exhibit pronounced two-fold oscillations. These angular dependences are well captured by
\begin{eqnarray}
R_{\parallel}(\phi) &=& R_0 - \Delta R \cos 2(\phi - \alpha), \\
R_{\perp}(\phi) &=& R_H + \Delta R \sin 2(\phi - \alpha),
\end{eqnarray}
where $\Delta R$ is the oscillation amplitude, $R_0$ is the average value of \Rpara$(\phi)$, $R_H$ is the average value of \Rperp$(\phi)$, and $\alpha$ denotes the principal axis orientation, defined by the direction of highest and lowest resistance.  

For disk-shaped samples, theoretical analysis~\cite{Vafek2023anisotropy} shows that fitting the angle-resolved transport to Eqs.~1--2 enables full reconstruction of the conductivity matrix. In particular, the ratio \DRR\ quantifies the strength of transport anisotropy, while a non-zero $R_H$ reflects antisymmetric off-diagonal components of the conductivity tensor, which corresponds to an anomalous Hall coefficient arising from orbital ferromagnetism.

As further support, Figs.~\ref{fig1}d--e show the magnetic-field dependence of $R_0$ and $R_H$, extracted from the angle-resolved transport measurements as the field $B$ is swept back and forth. The evolution of $R_H(B)$ exhibits a clear hysteresis loop, a hallmark of orbital ferromagnetism. The inset of Fig.~\ref{fig1}e displays the angular dependence measured at $B = 0$ for forward and backward sweeps; the change in the angular oscillation induced by sweeping $B$ points to a potential intertwinement between rotational and time-reversal symmetry breaking.

Within the triangular regime, shown in the zoomed-in map of Fig.~\ref{fig2}a, orbital ferromagnetism is tunable by the perpendicular electric field. Figures~\ref{fig2}b–c plot the anomalous Hall coefficient ($R_H$), with red and blue indicating opposite Hall responses, measured as the displacement field $D$ is swept back and forth. For both sweep directions, $R_H$ exhibits a sign reversal across $D = 0$. Moreover, changing the sweep direction of $D$ induces an additional sign reversal in $R_H$. This distinctive electric-field-driven switching generates a characteristic butterfly pattern in $R_H$, as illustrated in the line trace shown in Fig.~\ref{fig2}d. Together, the electric-field-driven switching of the Hall coefficient points to the emergence of multiferroicity  ~\cite{Han2023multiferroic}.

Most remarkably, the electric-field-driven transition in magnetism coincides with changes in transport anisotropy. As shown in Figs.~\ref{fig2}d--g, the butterfly-shaped line trace of $R_H$ is accompanied by hysteretic behavior in both $\Delta R / R_0$ and $\alpha$. This offers strong indication that the underlying rotational symmetry breaking is likewise tunable via the sweeping electric field.

This tunability is further demonstrated by comparing the angular dependence at 
$D = 25~\mathrm{mV/nm}$ (Fig.~\ref{fig2}h--i), corresponding to the red and blue open circles in Fig.~\ref{fig2}d--g. 
When $D$ is swept from positive to negative, the transport response exhibits an enhanced angular oscillation, 
indicating strong anisotropy, accompanied by a positive $R_H$ (Fig.~\ref{fig2}h). 
In contrast, sweeping $D$ from negative to positive results in a markedly reduced $\Delta R / R_0$, 
signaling suppressed anisotropy, while a negative $R_H$ reflects a $D$-induced reversal of the anomalous Hall effect. 
Furthermore, reversing the direction of the $D$ sweep rotates the principal axis by approximately $30^{\circ}$ 
(see Fig.~\ref{Dswitch} for electric-field--driven switching at negative $D$).

The hysteretic switching observed in the angle-resolved transport demonstrates an unprecedented level of control over the electronic order responsible for rotational symmetry breaking. Unlike strain- or distortion-induced effects from the underlying lattice, whose orientation is fixed, the demonstrated switching indicates that the observed anisotropy is fundamentally electronic in origin.

To further investigate the nature of the emerging order, we analyze the temperature-dependent onset of anisotropy and anomalous Hall effect. Fig.~\ref{fig3}a-b plot $|R_H|$ and \DRR\ as a function of $T$, revealing three distinct temperature regimes, denoted as regimes (i), (ii), and (iii). Transport signatures of each regime is discussed as follow.

\begin{itemize}
    \item Regime (i): At elevated temperature, angle-resolved transport exhibits a largely isotropic response with vanishing Hall coefficient (Fig.~\ref{fig3}c and f). 
    \item Regime (ii): As temperature decreases, an anomalous Hall effect emerges with a sharp onset in $|R_H|$ around $T = 4$ K (Fig.~\ref{fig3}a), accompanied by an enhancement in transport anisotropy (Fig.~\ref{fig3}b and d). Notably, electric-field-driven switching is absent in this regime (Fig.~\ref{fig3}g). 
   \item Regime (iii): the butterfly-shaped hysteresis in $R_H$ arises below $T = 1.5$ K  (Fig.~\ref{fig3}h), signaling the emergence of multiferroicity. This coincides with a second sharp increase in transport anisotropy (Fig.~\ref{fig3}b).
\end{itemize}

This temperature dependence allows us to make several significant observations. 
First, the onset of anisotropy is well described by a Curie--Weiss form (black dashed line in the inset of Fig.~\ref{fig3}b), 
which diverges near $T \approx 2.1~\mathrm{K}$. 
This behavior closely resembles nematic transitions observed in other quantum materials, 
such as iron-based superconductors~\cite{Chu2012divergent,Fernandes2014nematicity,Bohmer2022nematicity}. 
Following the convention established in prior works, we identify the cusp in \DRR\ as the nematic instability 
and label the corresponding transition temperature as $T_{\mathrm{nem}}$. 
This Curie--Weiss divergence marks the first nematic transition identified in a two-dimensional material.

Second, the absence of electric-field-driven switching in regime (ii) suggests that the ferromagnetic and ferroelectric responses originate from distinct mechanisms. 
We therefore associate the second onset in \DRR, which coincides with the appearance of $D$-driven switching, with the onset of multiferroicity, defining the corresponding temperature as $T_{\mathrm{multi}}$.

Finally, the emergence of the anomalous Hall effect around $T_{\mathrm{nem}}$ indicates a common underlying mechanism for rotational and time-reversal symmetry breaking. 
This observation is consistent with theoretical predictions of Coulomb-driven instabilities in momentum space, 
where a collective flocking effect induces spontaneous momentum condensation~\cite{Dong2021momentum,Jung2015momentum,Huang2023momentum,Mandal2023valleynematic,Parra2025nematicQM}.

Beyond the coexistence of anisotropy and the anomalous Hall effect, theoretical work has suggested that momentum polarization may also manifest through the nonlinear Hall effect~\cite{Huang2023momentum}. 
In the following, we identify the nonlinear Hall response using a recently established framework, 
employing angle-resolved transport measurements in the nonlinear regime~\cite{Chichinadze2024nonlinearHall}.

To extract the nonlinear transport response, we apply an a.c. current bias at frequency $\omega$ and examine the second harmonic voltage response at frequency $2\omega$ ~\cite{Ma2019nonlinear,Kang2019nonlinear,He2022nonlinear,Zhang2024nonreciprocity,Chichinadze2024nonlinearHall}. A small a.c. bias of $10 - 15$ nA is used to ensure measurements remain within the quadratic response regime.

Fig.~\ref{fig4}a--b show the angular dependence of the nonlinear response measured in the multiferroic regime. 
Instead of the two-fold angular oscillations observed in the linear response (Fig.~\ref{fig1}c), the nonlinear signals in both the parallel ($V_{\parallel}^{2\omega}$) and perpendicular ($V_{\perp}^{2\omega}$) channels exhibit an angular periodicity of $2\pi$. 
Moreover, the two channels display a relative phase shift of $\pi/2$, indicating that the underlying order possesses only a single mirror axis~\cite{Zhang2024nonreciprocity}, marked by the dashed lines in Fig.~\ref{fig4}a--b. 
This mirror axis aligns with the principal direction of minimum resistivity in the linear transport anisotropy, suggesting a common origin underlying the angular dependence of both the linear and nonlinear responses.

Most importantly, the transverse nonlinear response displays a substantially larger oscillation amplitude, denoted 
$\Delta V_{\perp}^{2\omega}$, compared to the longitudinal channel. 
This behavior is in excellent agreement with the predicted angular dependence of the nonlinear Hall effect~\cite{Chichinadze2024nonlinearHall}, 
indicating that the nonlinear Hall coefficient constitutes the dominant contribution to the conductivity tensor in the nonlinear regime.

As shown in Fig.~\ref{fig4}d, the temperature dependence of the nonlinear Hall effect, characterized by 
$\Delta V_{\perp}^{2\omega}$, exhibits an onset around $T = 3.6~\mathrm{K}$, coinciding precisely with the emergence of 
the anomalous Hall effect. 
Furthermore, at fixed temperature and carrier density, the $D$ dependence of the nonlinear Hall response mirrors that of the 
anomalous Hall coefficient, forming two distinct peaks away from $D = 0$ (Fig.~\ref{fig4}e). 
Taken together, these observations reveal a direct link between transport anisotropy, orbital ferromagnetism, 
and the nonlinear Hall effect. This is in excellent agreement with theoretical predictions of momentum-space condensation~\cite{Dong2021momentum,Huang2023momentum}.

While the connection between the ferroelectric order and the coexisting momentum condensation remains an open question, 
we note that electric-field--driven switching is also observed in the nonlinear Hall effect. 
Fig.~\ref{fig4}f--g show the angular dependence of $V_{\perp}^{2\omega}$, measured at fixed carrier density ($n$) and 
displacement field ($D$). 
When $D$ is swept in opposite directions, the angular behavior of the nonlinear response exhibits pronounced changes in both 
the oscillation amplitude and the orientation of the underlying mirror axis. 
This tunability further demonstrates that strain and lattice distortion do not play a dominant role in generating the observed 
angular dependence. Instead, the rotational symmetry breaking originates from the electronic order itself.

Overall, we uncover compelling evidence for momentum-space condensation through the simultaneous breaking of 
rotational, time-reversal, and inversion symmetries~\cite{Huang2023momentum,Jung2015momentum,Dong2021momentum,Parra2025nematicQM}. 
By establishing an intrinsic link between orbital ferromagnetism and rotational symmetry breaking, 
momentum condensation emerges as a new form of correlated electronic order. 
More broadly, this mechanism provides a unifying framework for understanding a wide class of emergent phases in flat-band systems 
and may be key to elucidating the nature of unconventional superconductivity~\cite{Han2024chiral,Morissette2025stripe}. 
Furthermore, the intimate connection between distinct symmetry-breaking orders suggests new pathways for electrically controlling 
intertwined electronic phases in two-dimensional quantum materials.

\section*{acknowledgments}
J.I.A.L. wishes to acknowledge helpful discussion with Chunli Huang and Rafael Fernandes. E.M. and J.I.A.L. acknowledge support from U.S. National Science Foundation under Award DMR-2143384.  J.I.A.L. acknowledge partial support from the Air Force Office of Scientific Research.   K.W. and T.T. acknowledge support from the JSPS KAKENHI (Grant Numbers 21H05233 and 23H02052) and World Premier International Research Center Initiative (WPI), MEXT, Japan. 
Part of this work was enabled by the use of pyscan (github.com/sandialabs/pyscan), scientific measurement software made available by the Center for Integrated Nanotechnologies, an Office of Science User Facility operated for the U.S. Department of Energy.

\bibliography{Li_ref}

@article{Sharpe2019,
  title={Emergent ferromagnetism near three-quarters filling in twisted bilayer graphene},
  author={Sharpe, Aaron L and Fox, Eli J and Barnard, Arthur W and Finney, Joe and Watanabe, Kenji and Taniguchi, Takashi and Kastner, MA and Goldhaber-Gordon, David},
  journal={arXiv preprint arXiv:1901.03520},
  year={2019}
}

@article{Serlin2019,
  title={Intrinsic quantized anomalous Hall effect in a moir$\backslash$'e heterostructure},
  author={Serlin, M and Tschirhart, CL and Polshyn, H and Zhang, Y and Zhu, J and Watanabe, K and Taniguchi, T and Balents, L and Young, AF},
  journal={arXiv preprint arXiv:1907.00261},
  year={2019}
}

@article{Cao2020nematicity,
  title={Nematicity and Competing Orders in Superconducting Magic-Angle Graphene},
  author={Cao, Yuan and Rodan-Legrain, Daniel and Park, Jeong Min and Yuan, Fanqi Noah and Watanabe, Kenji and Taniguchi, Takashi and Fernandes, Rafael M and Fu, Liang and Jarillo-Herrero, Pablo},
  journal={arXiv preprint arXiv:2004.04148},
  year={2020}
}

@article{Wu2017nematic,
  title={Spontaneous breaking of rotational symmetry in copper oxide superconductors},
  author={Wu, J and Bollinger, AT and He, X and Bo{\v{z}}ovi{\'c}, I},
  journal={Nature},
  volume={547},
  number={7664},
  pages={432--435},
  year={2017},
  publisher={Nature Publishing Group}
}

@article{Wu2020nematic,
  title={Electronic nematicity in Sr2RuO4},
  author={Wu, Jie and Nair, Hari P and Bollinger, Anthony T and He, Xi and Robinson, Ian and Schreiber, Nathaniel J and Shen, Kyle M and Schlom, Darrell G and Bo{\v{z}}ovi{\'c}, Ivan},
  journal={Proceedings of the National Academy of Sciences},
  volume={117},
  number={20},
  pages={10654--10659},
  year={2020},
  publisher={National Acad Sciences}
}

@article{Chen2020ABC,
  title={Tunable correlated chern insulator and ferromagnetism in a moir{\'e} superlattice},
  author={Chen, Guorui and Sharpe, Aaron L and Fox, Eli J and Zhang, Ya-Hui and Wang, Shaoxin and Jiang, Lili and Lyu, Bosai and Li, Hongyuan and Watanabe, Kenji and Taniguchi, Takashi and others},
  journal={Nature},
  volume={579},
  number={7797},
  pages={56--61},
  year={2020},
  publisher={Nature Publishing Group}
}

@article{Lin2021SOC,
  title={Spin-orbit--driven ferromagnetism at half moir{\'e} filling in magic-angle twisted bilayer graphene},
  author={Lin, Jiang-Xiazi and Zhang, Ya-Hui and Morissette, Erin and Wang, Zhi and Liu, Song and Rhodes, Daniel and Watanabe, K and Taniguchi, T and Hone, James and Li, JIA},
  journal={Science},
  volume={375},
  issue={6579},
  pages={437--441},
  year={2022},
  publisher={American Association for the Advancement of Science}
}

@article{Lin2022SDE,
  title={Zero-field superconducting diode effect in small-twist-angle trilayer graphene},
  author={Lin, Jiang-Xiazi and Siriviboon, Phum and Scammell, Harley D and Liu, Song and Rhodes, Daniel and Watanabe, K and Taniguchi, T and Hone, James and Scheurer, Mathias S and Li, JIA},
  journal={Nature Physics},
  volume={18},
  number={10},
  pages={1221--1227},
  year={2022},
  publisher={Nature Publishing Group UK London}
}

@article{Vafek2023anisotropy,
  title = {Anisotropic resistivity tensor from disk geometry magnetoconductance},
  author = {Vafek, Oskar},
  journal = {Phys. Rev. Appl.},
  volume = {20},
  issue = {6},
  pages = {064008},
  numpages = {8},
  year = {2023},
  month = {Dec},
  publisher = {American Physical Society},
  doi = {10.1103/PhysRevApplied.20.064008},
  url = {https://link.aps.org/doi/10.1103/PhysRevApplied.20.064008}
}

@article{Fernandes2014nematicity,
  title={What drives nematic order in iron-based superconductors?},
  author={Fernandes, RM and Chubukov, AV and Schmalian, J},
  journal={Nature physics},
  volume={10},
  number={2},
  pages={97--104},
  year={2014},
  publisher={Nature Publishing Group UK London}
}

@article{Bohmer2022nematicity,
  title={Nematicity and nematic fluctuations in iron-based superconductors},
  author={B{\"o}hmer, Anna E and Chu, Jiun-Haw and Lederer, Samuel and Yi, Ming},
  journal={Nature Physics},
  volume={18},
  number={12},
  pages={1412--1419},
  year={2022},
  publisher={Nature Publishing Group UK London}
}

@Article{Zhang2024nonreciprocity,
author={Zhang, Naiyuan James
and Lin, Jiang-Xiazi
and Chichinadze, Dmitry V.
and Wang, Yibang
and Watanabe, Kenji
and Taniguchi, Takashi
and Fu, Liang
and Li, J. I. A.},
title={Angle-resolved transport non-reciprocity and spontaneous symmetry breaking in twisted trilayer graphene},
journal={Nature Materials},
year={2024},
month={Mar},
day={01},
volume={23},
number={3},
pages={356-362},
issn={1476-4660},
doi={10.1038/s41563-024-01809-z},
url={https://doi.org/10.1038/s41563-024-01809-z}
}

@article{Chichinadze2024nonlinearHall,
  title={Observation of giant nonlinear Hall conductivity in Bernal bilayer graphene},
  author={Chichinadze, Dmitry V and Zhang, Naiyuan James and Lin, Jiang-Xiazi and Wang, Xiaoyu and Watanabe, Kenji and Taniguchi, Takashi and Vafek, Oskar and Li, JIA},
  journal={arXiv preprint arXiv:2411.11156},
  year={2024}
}

@article{Jung2015momentum,
  title={Persistent current states in bilayer graphene},
  author={Jung, Jeil and Polini, Marco and MacDonald, Allan H},
  journal={Physical Review B},
  volume={91},
  number={15},
  pages={155423},
  year={2015},
  publisher={APS}
}

@article{Dong2021momentum,
  title = {Isospin- and momentum-polarized orders in bilayer graphene},
  author = {Dong, Zhiyu and Davydova, Margarita and Ogunnaike, Olumakinde and Levitov, Leonid},
  journal = {Phys. Rev. B},
  volume = {107},
  issue = {7},
  pages = {075108},
  numpages = {10},
  year = {2023},
  month = {Feb},
  publisher = {American Physical Society},
  doi = {10.1103/PhysRevB.107.075108},
  url = {https://link.aps.org/doi/10.1103/PhysRevB.107.075108}
}

@article{Huang2023momentum,
  title={Spin and orbital metallic magnetism in rhombohedral trilayer graphene},
  author={Huang, Chunli and Wolf, Tobias MR and Qin, Wei and Wei, Nemin and Blinov, Igor V and MacDonald, Allan H},
  journal={Physical Review B},
  volume={107},
  number={12},
  pages={L121405},
  year={2023},
  publisher={APS}
}

@article{Han2023multiferroic,
  title={Orbital multiferroicity in pentalayer rhombohedral graphene},
  author={Han, Tonghang and Lu, Zhengguang and Scuri, Giovanni and Sung, Jiho and Wang, Jue and Han, Tianyi and Watanabe, Kenji and Taniguchi, Takashi and Fu, Liang and Park, Hongkun and others},
  journal={Nature},
  volume={623},
  number={7985},
  pages={41--47},
  year={2023},
  publisher={Nature Publishing Group UK London}
}

@article{Chu2012divergent,
  title={Divergent nematic susceptibility in an iron arsenide superconductor},
  author={Chu, Jiun-Haw and Kuo, Hsueh-Hui and Analytis, James G and Fisher, Ian R},
  journal={Science},
  volume={337},
  number={6095},
  pages={710--712},
  year={2012},
  publisher={American Association for the Advancement of Science}
}

@article{Mandal2023valleynematic,
  title={Valley-polarized nematic order in twisted moir{\'e} systems: In-plane orbital magnetism and crossover from non-Fermi liquid to Fermi liquid},
  author={Mandal, Ipsita and Fernandes, Rafael M},
  journal={Physical Review B},
  volume={107},
  number={12},
  pages={125142},
  year={2023},
  publisher={APS}
}

@article{Das2024multiferroic,
  title={Superpolarized Electron-Hole Liquid and Multiferroicity in Multilayer Graphene},
  author={Das, Mainak and Huang, Chunli},
  journal={arXiv preprint arXiv:2404.10069},
  year={2024}
}

@article{Sodemann2015nonlinear,
  title={Quantum nonlinear Hall effect induced by Berry curvature dipole in time-reversal invariant materials},
  author={Sodemann, Inti and Fu, Liang},
  journal={Physical review letters},
  volume={115},
  number={21},
  pages={216806},
  year={2015},
  publisher={APS}
}

@article{Kang2019nonlinear,
  title={Nonlinear anomalous Hall effect in few-layer WTe2},
  author={Kang, Kaifei and Li, Tingxin and Sohn, Egon and Shan, Jie and Mak, Kin Fai},
  journal={Nature materials},
  volume={18},
  number={4},
  pages={324--328},
  year={2019},
  publisher={Nature Publishing Group}
}

@article{Ma2019nonlinear,
  title={Observation of the nonlinear Hall effect under time-reversal-symmetric conditions},
  author={Ma, Qiong and Xu, Su-Yang and Shen, Huitao and MacNeill, David and Fatemi, Valla and Chang, Tay-Rong and Mier Valdivia, Andr{\'e}s M and Wu, Sanfeng and Du, Zongzheng and Hsu, Chuang-Han and others},
  journal={Nature},
  volume={565},
  number={7739},
  pages={337--342},
  year={2019},
  publisher={Nature Publishing Group}
}

@article{He2022nonlinear,
  title={Graphene moir{\'e} superlattices with giant quantum nonlinearity of chiral Bloch electrons},
  author={He, Pan and Koon, Gavin Kok Wai and Isobe, Hiroki and Tan, Jun You and Hu, Junxiong and Neto, Antonio H Castro and Fu, Liang and Yang, Hyunsoo},
  journal={Nature Nanotechnology},
  volume={17},
  number={4},
  pages={378--383},
  year={2022},
  publisher={Nature Publishing Group}
}

@article{Calvera2024ValleyNematic,
  title={Theory of Coulomb driven nematicity in a multi-valley two-dimensional electron gas},
  author={Calvera, Vladimir and Valenti, Agnes and Huber, Sebastian D and Berg, Erez and Kivelson, Steven A},
  journal={arXiv preprint arXiv:2406.12825},
  year={2024}
}

@article{Ren2021NematicDecoupling,
  title={Nanoscale decoupling of electronic nematicity and structural anisotropy in FeSe thin films},
  author={Ren, Zheng and Li, Hong and Zhao, He and Sharma, Shrinkhala and Wang, Ziqiang and Zeljkovic, Ilija},
  journal={Nature Communications},
  volume={12},
  number={1},
  pages={10},
  year={2021},
  publisher={Nature Publishing Group UK London}
}

@article{Han2024RPG,
  title={Correlated insulator and Chern insulators in pentalayer rhombohedral-stacked graphene},
  author={Han, Tonghang and Lu, Zhengguang and Scuri, Giovanni and Sung, Jiho and Wang, Jue and Han, Tianyi and Watanabe, Kenji and Taniguchi, Takashi and Park, Hongkun and Ju, Long},
  journal={Nature Nanotechnology},
  volume={19},
  number={2},
  pages={181--187},
  year={2024},
  publisher={Nature Publishing Group UK London}
}

@article{Han2024chiral,
  title={Signatures of chiral superconductivity in rhombohedral graphene},
  author={Han, Tonghang and Lu, Zhengguang and Hadjri, Zach and Shi, Lihan and Wu, Zhenghan and Xu, Wei and Yao, Yuxuan and Cotten, Armel A and Sedeh, Omid Sharifi and Weldeyesus, Henok and others},
  journal={arXiv preprint arXiv:2408.15233},
  year={2024}
}

@article{Lu2024FCI,
  title={Fractional quantum anomalous Hall effect in multilayer graphene},
  author={Lu, Zhengguang and Han, Tonghang and Yao, Yuxuan and Reddy, Aidan P and Yang, Jixiang and Seo, Junseok and Watanabe, Kenji and Taniguchi, Takashi and Fu, Liang and Ju, Long},
  journal={Nature},
  volume={626},
  number={8000},
  pages={759--764},
  year={2024},
  publisher={Nature Publishing Group UK London}
}

@article{Zhou2021RTG,
  title={Half-and quarter-metals in rhombohedral trilayer graphene},
  author={Zhou, Haoxin and Xie, Tian and Ghazaryan, Areg and Holder, Tobias and Ehrets, James R and Spanton, Eric M and Taniguchi, Takashi and Watanabe, Kenji and Berg, Erez and Serbyn, Maksym and others},
  journal={Nature},
  volume={598},
  number={7881},
  pages={429--433},
  year={2021},
  publisher={Nature Publishing Group UK London}
}

@article{Holleis2025nematicity,
  title={Nematicity and orbital depairing in superconducting Bernal bilayer graphene},
  author={Holleis, Ludwig and Patterson, Caitlin L and Zhang, Yiran and Vituri, Yaar and Yoo, Heun Mo and Zhou, Haoxin and Taniguchi, Takashi and Watanabe, Kenji and Berg, Erez and Nadj-Perge, Stevan and others},
  journal={Nature Physics},
  pages={1--7},
  year={2025},
  publisher={Nature Publishing Group UK London}
}

@article{Parra2025nematicQM,
  title={Band Renormalization, Quarter Metals, and Chiral Superconductivity in Rhombohedral Tetralayer Graphene},
  author={Parra-Martinez, Guillermo and Jimeno-Pozo, Alejandro and Phong, Vo Tien and Sainz-Cruz, Hector and Kaplan, Daniel and Emanuel, Peleg and Oreg, Yuval and Pantaleon, Pierre A and Silva-Guillen, Jose Angel and Guinea, Francisco},
  journal={arXiv preprint arXiv:2502.19474},
  year={2025}
}

@article{Sha2024R4G,
  title={Observation of a Chern insulator in crystalline ABCA-tetralayer graphene with spin-orbit coupling},
  author={Sha, Yating and Zheng, Jian and Liu, Kai and Du, Hong and Watanabe, Kenji and Taniguchi, Takashi and Jia, Jinfeng and Shi, Zhiwen and Zhong, Ruidan and Chen, Guorui},
  journal={Science},
  volume={384},
  number={6694},
  pages={414--419},
  year={2024},
  publisher={American Association for the Advancement of Science}
}

@article{Liu2024R4G,
  title={Spontaneous broken-symmetry insulator and metals in tetralayer rhombohedral graphene},
  author={Liu, Kai and Zheng, Jian and Sha, Yating and Lyu, Bosai and Li, Fengping and Park, Youngju and Ren, Yulu and Watanabe, Kenji and Taniguchi, Takashi and Jia, Jinfeng and others},
  journal={Nature nanotechnology},
  volume={19},
  number={2},
  pages={188--195},
  year={2024},
  publisher={Nature Publishing Group UK London}
}

@article{Han2024R4G,
  title={Large quantum anomalous Hall effect in spin-orbit proximitized rhombohedral graphene},
  author={Han, Tonghang and Lu, Zhengguang and Yao, Yuxuan and Yang, Jixiang and Seo, Junseok and Yoon, Chiho and Watanabe, Kenji and Taniguchi, Takashi and Fu, Liang and Zhang, Fan and others},
  journal={Science},
  volume={384},
  number={6696},
  pages={647--651},
  year={2024},
  publisher={American Association for the Advancement of Science}
}

@article{Choi2025R4G,
  title={Superconductivity and quantized anomalous Hall effect in rhombohedral graphene},
  author={Choi, Youngjoon and Choi, Ysun and Valentini, Marco and Patterson, Caitlin L and Holleis, Ludwig FW and Sheekey, Owen I and Stoyanov, Hari and Cheng, Xiang and Taniguchi, Takashi and Watanabe, Kenji and others},
  journal={Nature},
  pages={1--6},
  year={2025},
  publisher={Nature Publishing Group UK London}
}

@article{Zhang2025angular,
  title={Angular Interplay of Nematicity, Superconductivity, and Strange Metallicity in a Moir{\'e} Flat Band},
  author={Zhang, Naiyuan J and Nosov, Pavel A and Sommer, Ophelia Evelyn and Wang, Yibang and Watanabe, Kenji and Taniguchi, Takashi and Khalaf, Eslam and Li, JIA},
  journal={arXiv preprint arXiv:2503.15767},
  year={2025}
}

@article{Morissette2025stripe,
  title={Superconductivity, Anomalous Hall Effect, and Stripe Order in Rhombohedral Hexalayer Graphene},
  author={Morissette, Erin and Qin, Peiyu and Wu, Hai-Tian and Zhang, Naiyuan J and Watanabe, K and Taniguchi, T and Li, JIA},
  journal={arXiv preprint arXiv:2504.05129},
  year={2025}
}

\newpage
\clearpage

\section*{Method}

\renewcommand{\thefigure}{M\arabic{figure}}
\def\theequation{M\arabic{equation}} 
\def\thetable{M\Roman{table}}
\setcounter{figure}{0}
\setcounter{equation}{0}

In this section, we provide detailed discussions to further substantiate results reported in the main text. This section offers a comprehensive review, summarizing the notations employed. We also expand on the nuance associated with angle-resolved transport measurement in the presence of electric-field-driven switching. 


\subsection{I. Angle-resolved transport measurement}

\begin{figure}[!b]
\includegraphics[width=0.95\linewidth]{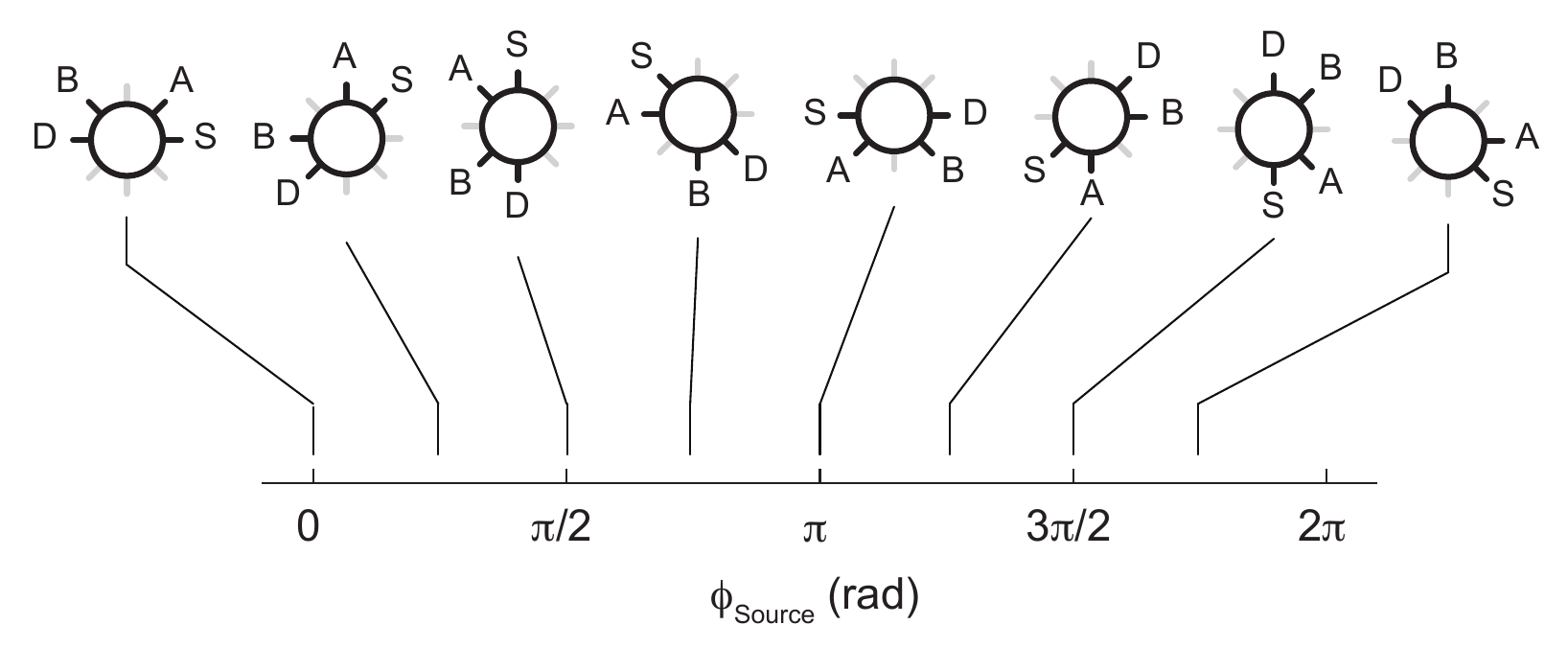}
\caption{\label{ARTM}{\bf{Definition of $\phi$.}} 
Schematic of the angle-resolved transport measurement setup used to extract $R_{\parallel}$ in a sample shaped into the so-called ``sunflower'' geometry.
}
\end{figure}

In the sunflower sample geometry, angle‐resolved transport is performed using specific measurement configurations, such as $R_{\parallel}$ and $R_{\perp}$, as illustrated in Fig.~\ref{fig1}b. Fig.~\ref{ARTM} shows a schematic of angle‐resolved transport for the $R_{\parallel}$ configuration. By rotating the measurement configuration, we probe the transport response for different current‐flow directions. We define $\phi = 0$ by the first configuration on the left, and each subsequent configuration to the right increases $\phi$ by $45^{\circ}$. The resulting angular dependence exhibits a sinusoidal oscillation with a period of $\pi$.

Eq.~1--2 follow from the theoretical framework proposed in Ref.~\cite{Vafek2023anisotropy}. 
By fitting the angular dependence of $R_{\parallel}$ and $R_{\perp}$ using Eqs.~1--2, 
we extract the conductivity tensor describing transport response in the linear regime, parameterized by the orientation of the principal axes $\alpha$ 
and the resistivities along each principal direction, $\rho_{xx}$ and $\rho_{yy}$. 
Throughout this work, \DRR\ quantifies the difference between $\rho_{xx}$ and $\rho_{yy}$ 
and serves as a measure of the strength of the transport anisotropy.

We note that angle-resolved transport is not limited to the configurations shown in Fig.~\ref{ARTM}. 
The sunflower geometry enables a broader set of measurement configurations, all of which can be used to extract the full 
conductivity matrix~\cite{Chichinadze2024nonlinearHall}. 
However, given the excellent quality of the angular fits obtained from $R_{\parallel}$ and $R_{\perp}$, these additional configurations are not required. 
Accordingly, in this work we restrict our analysis to the configurations illustrated in Fig.~\ref{fig1}b.

\begin{figure*}
\includegraphics[width=0.92\linewidth]{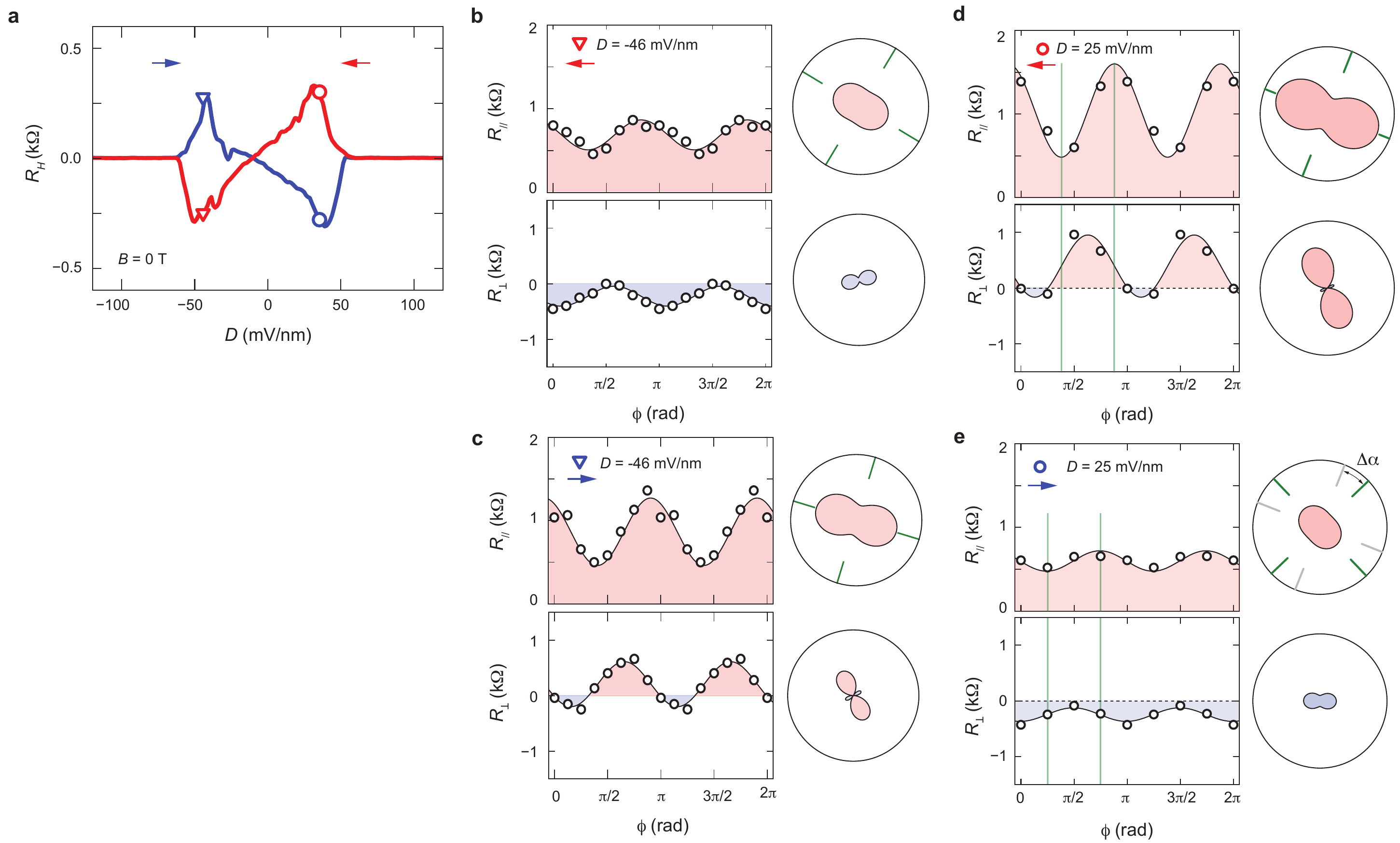}
\caption{\label{Dswitch}{\bf{Electric-field-driven switching of transport anisotropy.}} 
(a) Butterfly-shaped hysteresis in the anomalous Hall coefficient, extracted by fitting angular dependent transport response using Eq.~1-2. (b-c) Angle-resolved transport response measured at $D = -46$ mV/nm, as $D$ is swept (b) from positive to negative, and (c) from negative to positive (corresponding to the red and blue triangles in panel (a)). (d-e) Angle-resolved transport response measured at $D = +25$ mV/nm as $D$ is swept (d) from positive to negative, and (e) from negative to positive (corresponding to the red and blue circles in panel (a)).
}
\end{figure*}

\subsection{II. Electric- and magnetic-field-driven hysteresis}

Following the experimental report of multiferroic order in rhombohedral pentalayer graphene (R5G), theoretical models have 
been proposed to account for the electric-field--driven transitions in the ferromagnetic order~\cite{Das2024multiferroic}. 
These works argue that the butterfly-shaped hysteresis loop (Fig.~\ref{fig2}d and Fig.~\ref{Dswitch}a) arises from coupling 
between the magnetic order and a spontaneous layer polarization. 
Within this framework, $D$-driven hysteresis results from the system entering an out-of-equilibrium layer-polarized state. 
For example, when $D$ is swept from positive to negative, the anomalous Hall coefficient $R_H$ at positive $D$ 
($R_H > 0$) corresponds to the ground-state order (red circle in Fig.~\ref{Dswitch}a).  whereas the negative $R_H$ observed 
at negative $D$ reflects an out-of-equilibrium configuration (red triangle in Fig.~\ref{Dswitch}a).

Fig.~\ref{Dswitch}b--e show the angular-dependent transport response measured at two values of $D$ within the 
butterfly-shaped hysteresis loop. 
Panels~\ref{Dswitch}b and \ref{Dswitch}d correspond to measurements taken as $D$ is swept from positive to negative, 
whereas panels~\ref{Dswitch}c and \ref{Dswitch}e are obtained as $D$ is swept from negative to positive. 
According to the proposed model, panels~\ref{Dswitch}d and \ref{Dswitch}c represent equilibrium states, in which the 
combination of orbital ferromagnetism and layer polarization minimizes the system's energy. 
In contrast, panels~\ref{Dswitch}b and \ref{Dswitch}e correspond to out-of-equilibrium configurations: 
here, despite the sign reversal of the anomalous Hall coefficient $R_H$, the underlying magnetization remains unchanged 
until the multiferroic order collapses as $D$ is tuned outside the triangular regime.

Most interestingly, the equilibrium state exhibits markedly stronger anisotropy, with $\Delta R / R_0$ exceeding 0.5 
(Fig.~\ref{Dswitch}d and c). 
By contrast, the anisotropy is substantially suppressed in the out-of-equilibrium state (Fig.~\ref{Dswitch}b and e). 
This behavior demonstrates a distinctive evolution of the anisotropy strength across the butterfly-shaped hysteresis loop. 
Although the microscopic origin of this evolution remains an open question, it highlights an intrinsic connection between 
rotational symmetry breaking and the coexisting emergent orders.




\newpage

\newpage
\clearpage

\clearpage

\newpage
\begin{widetext}
\section{Supplementary Materials}

\begin{center}
\textbf{\large Evidence of Momentum Space Condensation in Rhombohedral Hexalayer Graphene}\\
\vspace{10pt}

Erin Morissette$^{\ast}$,
Peiyu Qin$^{\ast}$,
Kenji Watanabe, Takashi Taniguchi, and J.I.A. Li$^{\dag}$

\vspace{10pt}
$^{\dag}$ Corresponding author. Email: jia.li@austin.utexas.edu
\end{center}


\renewcommand{\vec}[1]{\boldsymbol{#1}}

\renewcommand{\thefigure}{S\arabic{figure}}
\def\theequation{S\arabic{equation}}
\def\thetable{S\Roman{table}}
\setcounter{figure}{0}
\setcounter{equation}{0}

\subsection{Fermiology and quantum oscillation}

We investigate the fermiology of the multiferroic state by analyzing the Shubnikov–de Haas (SdH) quantum oscillations. As shown in Fig.~\ref{fig5}a, quantum oscillations exhibit a ``fish-net''  pattern across a wide carrier density range. We perform a Fourier transform of $R_{\parallel}$ as a function of $1/B$ to identify the the Luttinger volume and degeneracy of the underlying Fermi sea (Fig.~\ref{fig5}b).

Figure~\ref{fig5}c marks the center frequencies of the most prominent peaks in the Fourier spectrum, revealing distinct density regimes characterized by different Fermi surface topologies. At low carrier density ($n = -0.5 \times 10^{12}$ cm$^{-2}$), the two dominant frequencies, $f_1$ and $f_2$, satisfy $f_1 - f_2 = 1$, suggesting two possible Fermi surface topologies: (i) a quarter metal characterized by an annular Fermi sea or (ii) an exotic half-metal, where an electron pocket coexists with a hole pocket, each featuring opposite valley polarizations (inset of Fig.~\ref{fig5}c). 

At higher densities ($-2.5 \times 10^{12}$ cm$^{-2} < n < -1.0 \times 10^{12}$ cm$^{-2}$), the sum rule $f_3 - f_4 = 1/2$ suggests either (i) a half-metal phase with an annular Fermi sea or (ii) the coexistence of two electron and two hole pockets, indicative of a more complex isospin texture. 

Since angle-resolved transport measurements reveal no anomalous Hall effect in the low- and high-density regimes (Fig.~\ref{fig2}b-c), the Fermi surface likely consists of coexisting electron and hole pockets with opposite valley polarizations, where the overall Chern number remains close to zero. This is in line with prior theoretical predictions~\cite{Das2024multiferroic}. In the presence of an out-of-plane electric field, the electron and hole pockets polarize to opposite graphene layers, creating a distinct charge distribution that likely serves as the direct link between transport anisotropy and ferroelectricity.

As the multiferroic order emerges during the transition between these high- and low-density regimes, the spectral weight of the Fourier transform shifts from \( f_1 \) and \( f_2 \) to \( f_3 \) and \( f_4 \), signaling the onset of a partially isospin-polarized (PIP) phase~\cite{Zhou2021RTG,Holleis2025nematicity}. This suggests that Fermi surface reconstruction plays a pivotal role in enabling momentum-space instabilities, which, in turn, are responsible for the simultaneous breaking of time-reversal, rotational, and inversion symmetries.

At very low density quantum oscillations reveal dominating frequency peaks at $f_v= 1/3$ and $1/6$ (Fig.~\ref{fan}), indicating the presence of three distinct Fermi pockets with a two-fold degeneracy.

\begin{figure*}
\includegraphics[width=0.99\linewidth]{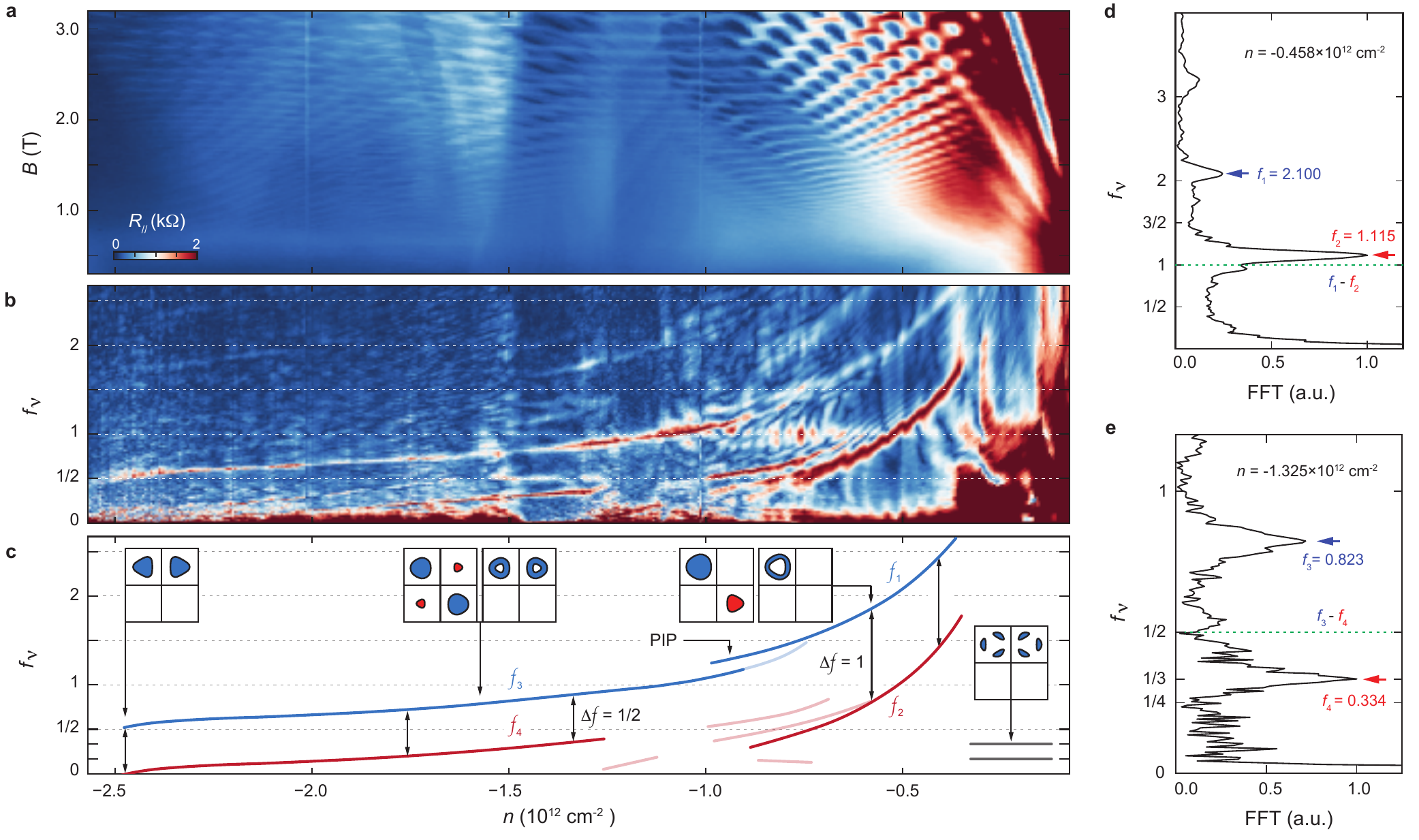}
\caption{\label{fig5}{\bf{Fermiology and coexistence of electron and hole fermi pockets.}} (a) \Rpara\ measured as a function of $n$ and $B$ measured at $D = 75$ mV/nm. (b) Fourier transform of quantum oscillation in (a). (c) Solid lines label the center frequencies of the most prominent peaks in \emph{f}$_\nu$ from (b). Inset shows possible fermi surface contours in different density regimes, extracted based on the intensity peaks in \emph{f}$_\nu$. (d-e) Fourier transform amplitude of data in (b) for $n = -0.458 \times 10^{12}$ cm$^{-2}$ (d) and $n = -1.325 \times 10^{12}$ cm$^{-2}$ (e).}
\end{figure*}

\begin{figure*}
\includegraphics[width=0.4\linewidth]{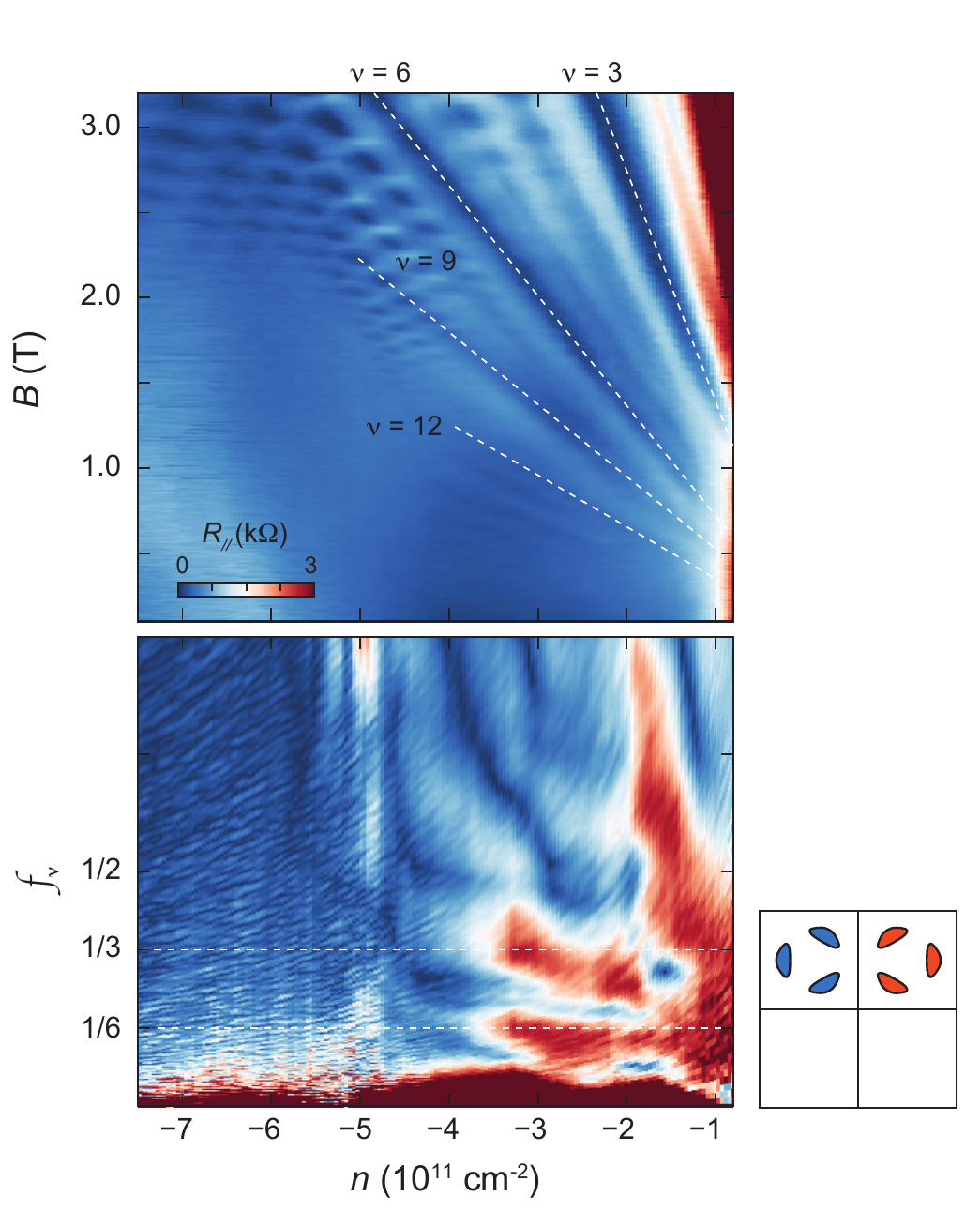}
\caption{\label{fan}{\bf{Fermiology at low carrier density.}} 
Longitudinal resistance measured as a function of carrier density $n$ and perpendicular magnetic field $B$ at 
$D = 75~\mathrm{mV/nm}$ (top), and the corresponding Fourier transform of the quantum oscillations (bottom). 
The measurement is performed at a density lower than that of the multiferroic regime. 
The two dominant frequencies, $f_{\nu} = 1/3$ and $1/6$, indicate the presence of three distinct Fermi pockets with a two-fold degeneracy.
}
\end{figure*}

\subsection{$B$ and $D$ induced hysteresis of multiferroicity}

\begin{figure*}
\includegraphics[width=0.95\linewidth]{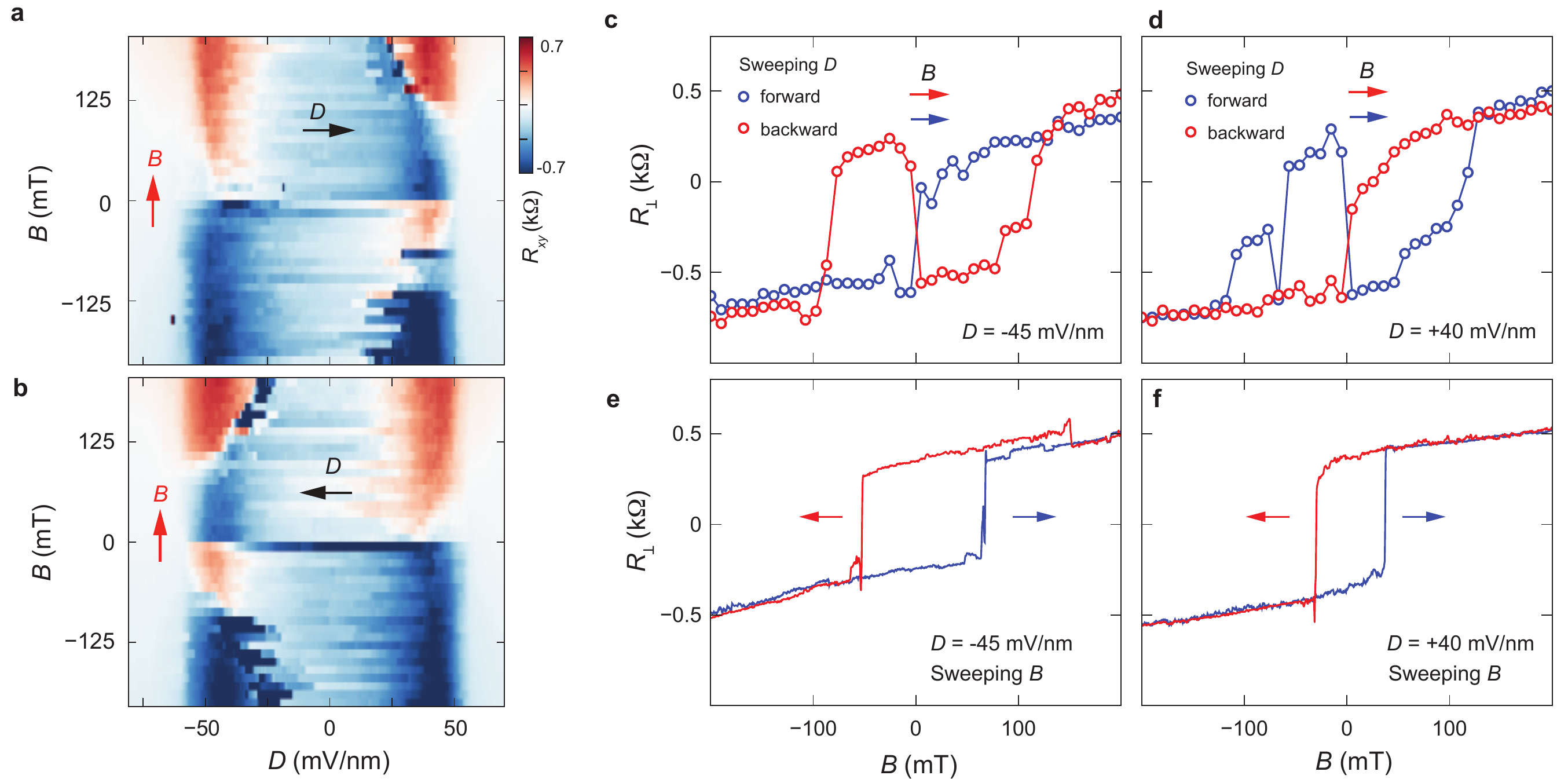}
\caption{\label{bhyst}{\bf{Electric- and magnetic-field-driven hysteresis.}} 
(a-b) Hall resistance sweeping displacement field $D$ (a) forwards and (b) backwards as the fast loop and sweeping magnetic field $B$ slowly from $-150$ mT to $+150$ mT.  
Vertical linecuts of these color maps are plotted in (c) for $D = -45$ mV/nm and (d) for $D = 40$ mV/nm as a function of magnetic field, with the data sweeping $D$ forward and backward shown in blue and red circles respectively. 
(e-f) Magnetic field hysteresis loops of Hall resistance \Rperp\ sweeping $B$ forward and backward (blue and red trace, respectively) for (a) $D = -45$ mV/nm and (b) $D = 40$ mV/nm. All measurements are performed at $T = 40$ mK, with the sample tuned to charge carrier density of $n = -0.82 \times 10^{12}$ cm$^{-2}$.
}
\end{figure*}

Within the multiferroic regime, sweeping electric  and magnetic-field produces qualitatively different hysteresis loops, as illustrated in Fig.~\ref{bhyst}a-b. When the displacement field $D$ is swept from negative to positive (or vice versa), the Hall coefficient measured at positive (negative) $D$ exhibits a sign reversal at small $B$.  

This behavior is further illustrated by the $B$-dependent line cuts shown in Fig.~\ref{bhyst}c-d, taken from Fig.~\ref{bhyst}a-b. Each point in these line cuts corresponds to a fixed $B$-field while $D$ is swept forward (blue circles) or backward (red circles). Throughout these measurements, the out-of-plane $B$-field is slowly swept from $-150$ mT to $+150$ mT.

By contrast, sweeping $B$ at a fixed $D$ yields a different hysteresis pattern (Fig.~\ref{bhyst}e-f), in which back-and-forth changes in $B$ produce identical loops at $D = -45$ and $+40$ mV/nm.  

These distinct responses arise because varying $B$ primarily reverses the orbital magnetic order, whereas varying $D$ switches the magnetism and the layer polarization simultaneously, activating different aspects of the multiferroc state  ~\cite{Das2024multiferroic}.

\begin{figure*}
\includegraphics[width=\linewidth]{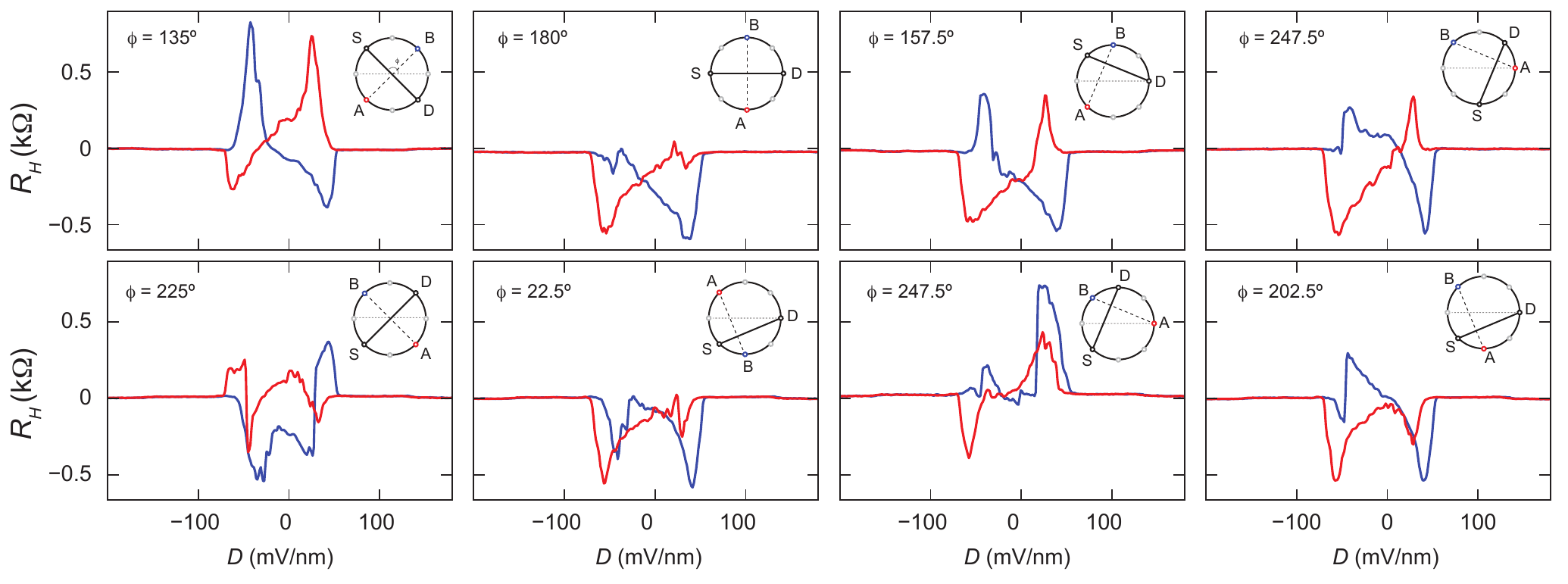}
\caption{\label{butterfly}{\bf{Electric-field-driven hysteresis in transport anisotropy.}} 
\Rperp\ measured as a function of $D$ with current flowing in different directions.
}
\end{figure*}

\begin{figure*}
\includegraphics[width=0.95\linewidth]{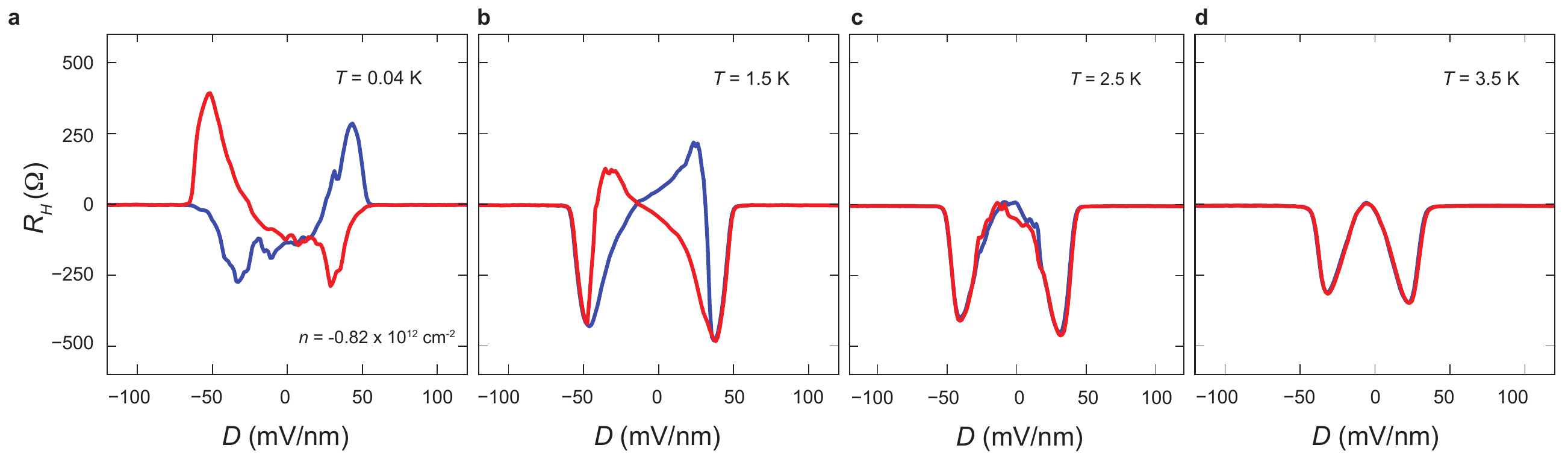}
\caption{\label{TD}{\bf{D-induced switching at different temperature.}} $R_H$ measured while $D$ is swept back and forth at (a) $T = 40$ mK, (b) $T = 1.5$ K, (c) $T = 2.5$ K, and (d) $T = 3.5$ K.
}
\end{figure*}

\subsection{Coupling between multiferroicity and transport anisotropy}

As shown in Fig.~\ref{fig2}, electric-field-driven transitions induce a pronounced rotation of the principal axes. Due to the evolution of anisotropy as a function of $D$, an accurate characterization of anomalous Hall coefficient as a function of $D$ cannot be extracted by measuring transverse resistance along a single direction.

To illustrate this, Fig.~\ref{butterfly} presents $R_{\perp}$ for various angles ($\phi$). The presence of transport anisotropy implies that $R_{\perp}$ measured along a fixed current direction does not directly correspond to the Hall coefficient ($R_H$). For instance, at $\phi = 180^\circ$, $R_{\perp}$ remains negative across the multiferroic regime, even though the underlying Hall coefficient undergoes a sign reversal at $D=0$. Additionally, for certain current orientations, such as $\phi = 225^{\circ}$, $D$-driven switching in $R_{\perp}$ appears random (Fig.~\ref{butterfly}), further emphasizing the role of anisotropy in governing transport responses.


\subsection{Temperature dependence on the butterfly-shaped hysteresis loop}

Figure~\ref{TD} plots $R_H$ at different temperatures, where $R_H$ is extracted from angle-resolved measurements while $D$ is swept back and forth. Near the boundary between regime (ii) and (iii), the butterfly-shaped hysteresis loop in $R_H$ gradually reduces in size, before fully disappearing at higher temperatures. In this intermediate regime, the hysteresis loop remains centered around $D = 0$ but is slightly diminished. At large $D$, the sign of $R_H$ reverts to its value in regime (ii), where the same $R_H$ is observed at positive and negative $D$.

These measurements are performed at $B=0$. The sign of $R_H$ at $B = 0$ can be switched by training with a positive magnetic field.

\subsection{Sample preparation}

Fig.~\ref{Sample}b shows the hexalayer graphene flake, imaged using the Kelvin probe microscopy. The color contrast reveals regimes with rhombohedral (dark color) and bernal-stacking (light color).

After AFM scanning, the rhombohedral regime is cut, using the lithography mode of the AFM. This separates the regimes with different stacking orders, preventing the domain boundary to relax during the stacking procedure. 

\begin{figure*}
\includegraphics[width=0.99\linewidth]{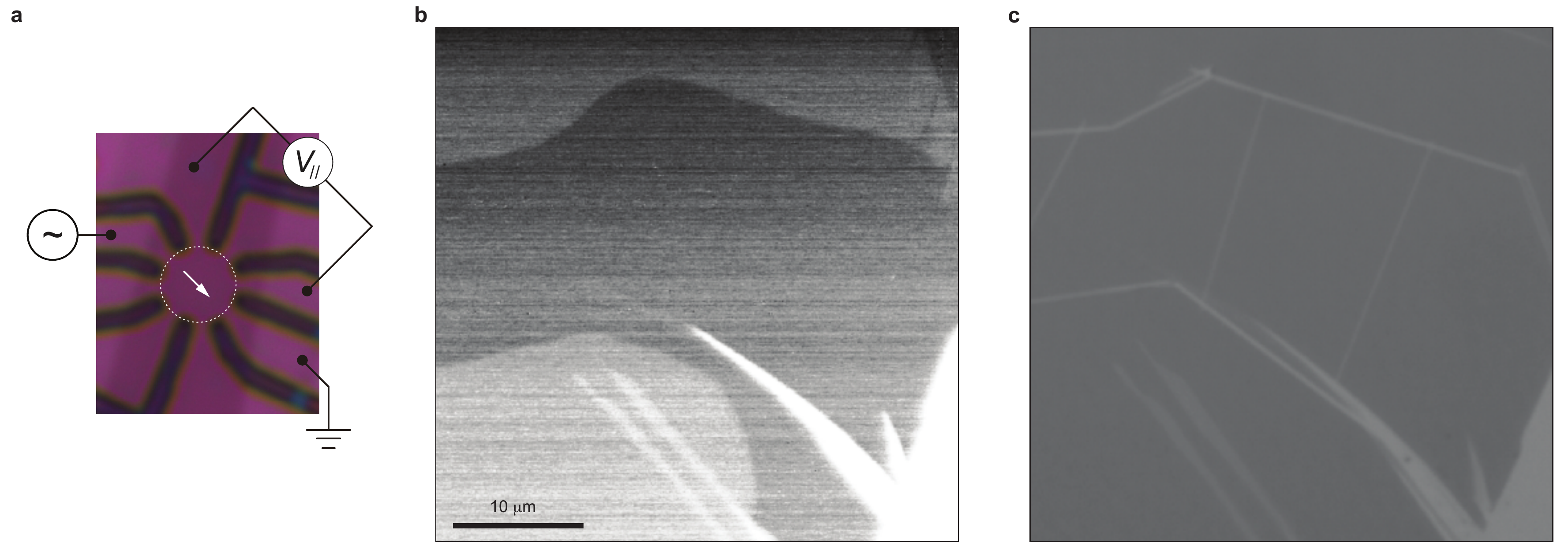}
\caption{\label{Sample}{\bf{Sample fabrication.}} (a) An optical image of the RHG sample patterned into the ``sunflower'' geometry.The diameter of the disk-shaped channel, highlighted by the white dashed line, is $3 \mu$m. (b) The hexalayer graphene flake is imaged using the Kelvin probe microscopy. The color contrast reveals regimes with rhombohedral (dark color) and bernal-stacking (light color). (c) The hexalayer graphene crystal is cut using the lithography mode of the AFM. This separates the regimes with different stacking orders, preventing the domain boundary to relax during the stacking procedure.  
}
\end{figure*}

\begin{figure*}
\includegraphics[width=0.65\linewidth]{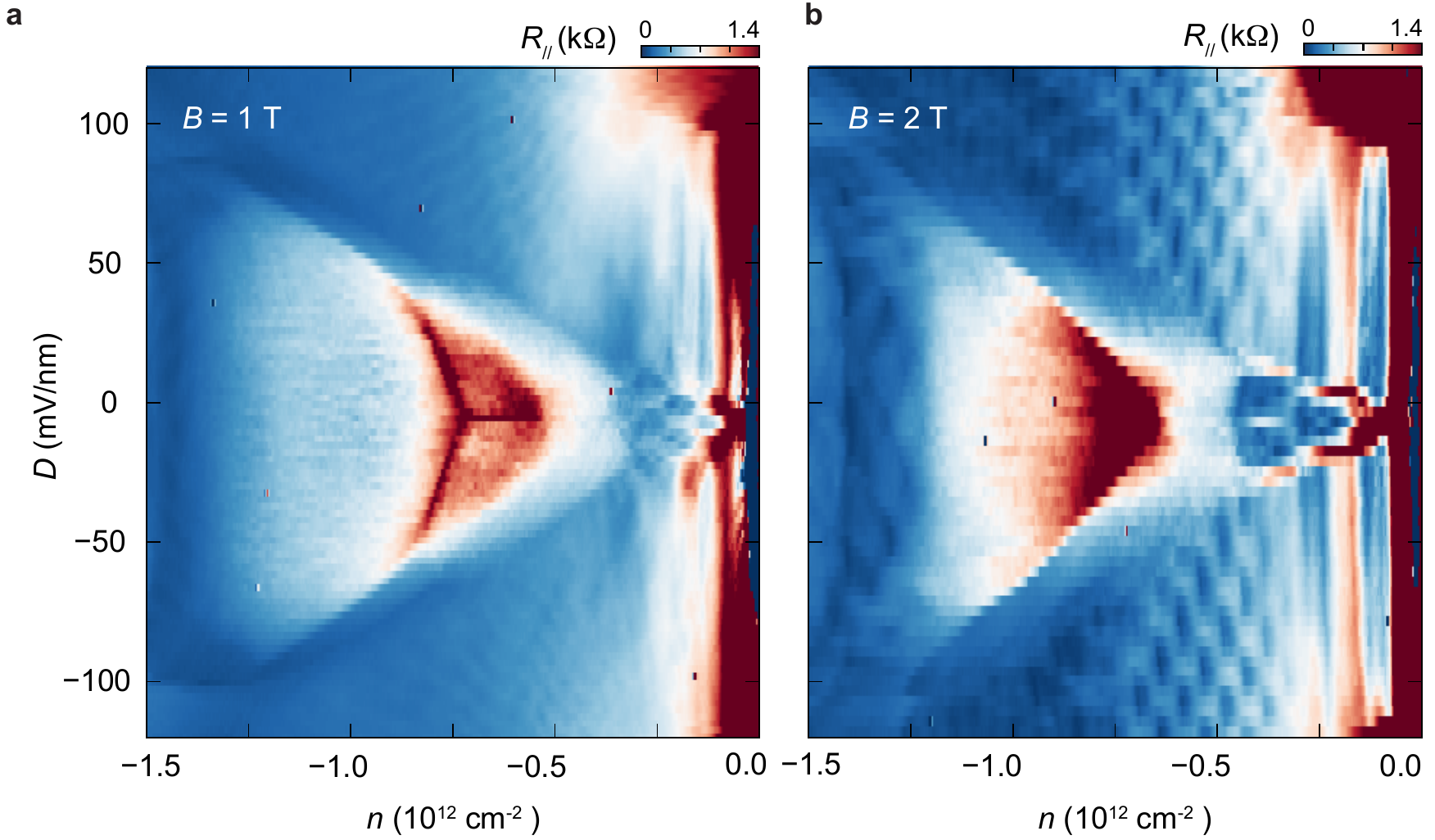}
\caption{\label{nDmap}{\bf{Multiferroic regime at finite magnetic field.}} Longitudinal resistance as a function of carrier density $n$ and displacement field $D$ in the multiferroic regime at (a) $B$ = 1 T and (b) 2 T. 
Shubnikov-de Haas (SdH) quantum oscillations are visible outside the multiferroic regime, forming the ``fish-net" pattern discussed in the main text and showcasing the three-fold Landau level sequence at low carrier density. 
}
\end{figure*}


\end{widetext}

\end{document}